\newif\ifarxiv

\documentclass[12pt]{article}
\newif\ifarxiv
\arxivtrue

\ifarxiv
\usepackage{arxiv}
\usepackage[authoryear,round,longnamesfirst]{natbib}
\usepackage{fancyvrb}
\usepackage{graphicx}

\usepackage{xcolor}
\colorlet{orangecolour}{green!10!orange!90!}
\RequirePackage{hyperref} 
\let\code=\texttt
\let\proglang=\textsf
\newcommand{\pkg}[1]{{\fontseries{m}\fontseries{b}\selectfont #1}}
\newcommand{\email}[1]{\href{mailto:#1}{\normalfont\texttt{#1}}}
\newcommand{\Plainauthor}[1]{\def\@Plainauthor{#1}}
\newcommand{\Plaintitle}[1]{\def\@Plaintitle{#1}}
\newcommand{\Abstract}[1]{\def\@Abstract{#1}}
\newcommand{\Keywords}[1]{\def\@Keywords{#1}}
\newcommand{\Plainkeywords}[1]{\def\@Plainkeywords{#1}}
\newcommand{\Address}[1]{\def\@Address{#1}}
\DefineVerbatimEnvironment{Code}{Verbatim}{}
\DefineVerbatimEnvironment{CodeInput}{Verbatim}{fontshape=sl}
\DefineVerbatimEnvironment{CodeOutput}{Verbatim}{}
\newenvironment{CodeChunk}{}{}
\AtBeginDocument{\maketitle\begin{abstract}
  The penalized Cox proportional hazard model is a popular analytical approach for survival data with a large number of covariates. Such problems are especially challenging when covariates vary over follow-up time (i.e., the covariates are time-dependent). The standard \proglang{R} packages for fully penalized Cox models cannot currently incorporate time-dependent covariates. To address this gap, we implement a variant of gradient descent algorithm (\emph{proximal gradient descent}) for fitting penalized Cox models. We apply our implementation to real and simulated data sets.

\end{abstract}}
\date{}
\fi

\newcommand{\pcoxtime}{\pkg{pcoxtime}\xspace}

\usepackage{lmodern}
\usepackage{amsmath}

\usepackage{framed}

\usepackage{xspace}
\usepackage{float}

\usepackage{xcolor}
\colorlet{orangecolour}{green!10!orange!90!}

\newcommand{\comment}{No comment command defined}
\renewcommand{\comment}[2]{} 
\renewcommand{\comment}[2]{{\color{blue} \emph{[[#1: #2]]}}}

\newcommand{\penL}{\ensuremath{\Omega}}

\newcommand{\CPH}{CPH\xspace}
\newcommand{\algorithm}[1]{\textbf{#1}}
\newcommand{\sstart}{\textrm{start}}
\newcommand{\sstop}{\textrm{stop}}
\newcommand{\mx}{\ensuremath{\max}}
\newcommand{\lammx}{\ensuremath{\lambda_{\mx}}}
\newcommand{\CVPL}{\widehat{\mathrm{\text{CV-PLD}}(\lambda)}}
\newcommand{\minCVPL}{Min.\ CV-PLD\xspace}

\author{Steve Cygu\\McMaster University
   \And Jonathan Dushoff\\McMaster University
   \And Benjamin M. Bolker\\McMaster University}
\Plainauthor{}

\title{\proglang{pcoxtime:} Penalized Cox Proportional Hazard Model for Time-dependent Covariates}
\Plaintitle{pcoxtime: Penalized Cox Proportional Hazard Model for Time-dependent Covariates}

\Abstract{
  The penalized Cox proportional hazard model is a popular analytical approach for survival data with a large number of covariates. Such problems are especially challenging when covariates vary over follow-up time (i.e., the covariates are time-dependent). The standard \proglang{R} packages for fully penalized Cox models cannot currently incorporate time-dependent covariates. To address this gap, we implement a variant of gradient descent algorithm (\emph{proximal gradient descent}) for fitting penalized Cox models. We apply our implementation to real and simulated data sets.

}

\Plainkeywords{survival, time-dependent, Cox proportional hazard, penalized, elastic net, proximal}

\Address{
  Steve Cygu, Jonathan Dushoff, Benjamin M. Bolker\\
  School of Computational Science and Engineering\\
  \emph{and}\\
  Department of Biology\\
  McMaster University\\
  1280 Main St W, Hamilton, ON L8S 4L8, Canada\\
  E-mail: \email{cygu@aims.ac.za}\\
}

\begin{document}


\section{Introduction} \label{sec:intro}


Survival analysis studies event times, such as time to cancer recurrence or time to death. Its goal is to predict the time-to-event (survival time) using a set of covariates, and to estimate the effect of different covariates on survival. Survival models typically attempt to estimate the \emph{hazard}, the probability (density) of the occurrence of the event of interest within a specific small time interval. Binary classification methods from machine learning can be used in problems that focus on predicting whether an event occurs within a specified time window. However, while binary classifiers can predict outcomes for a specified time window, they fail to account for one of the unique characteristics of survival data --- \emph{censoring}. In survival data, some of the subjects may be lost to follow-up, or may be event-free by the end of the follow-up time; hence the event times represent censoring times rather than failure (death, recurrence, etc.) times. Since binary classifiers consider only whether or not the event occurred in the last observation window, they lack the interpretability and flexibility of models that consider hazards as a function of time \citep{kvamme2019time}.

Cox proportional hazard (\CPH) models are the most common approach in survival analysis. Traditionally, the \CPH model has been applied in problems where the number of observations, $n$, is much larger than the number of covariates, $p$. In the modern era of big data, however, researchers often encounter cases where $p \approx n$ (or $p \gg n$). In cancer research, for example, rapid advances in genomic technologies have led to the  generation of vast amounts of cancer data \citep{cagan2017rethinking} --- presenting inherent challenges for effective and efficient data analysis. Penalized regression methods such as \emph{lasso}, \emph{ridge} or \emph{elastic net} offer a statistically convenient way of handling high-dimensional data, especially when building predictive models. The subclass of penalized methods which are sparsity-inducing (e.g. lasso and elastic net) can also be used to select useful predictive features from a large set.

The standard \CPH model (i.e., with no time-dependent covariates) assumes that the hazard ratio is constant over the entire follow-up period, or equivalently that each covariate is fixed over time and has a constant multiplicative effect on the hazard function. This assumption is problematic when covariates of interest themselves change over time. For example, cancer patients' healthcare access may change over the course of a study. Some implementations of \CPH models allow such \emph{time-dependent covariates}. However, their use requires more attention than the fixed (time-independent) covariates \citep{hochstein2013survival, therneau2017using, austin2020review}. 

Many authors have implemented \CPH models with penalization, but many implementations \citep{gui2005penalized, park2007l1, sohn2009gradient, goeman2010l1} are computationally inefficient, due to their use of the Newton-Raphson algorithm \citep{gorst2012coordinate}. Some newer implementations are more efficient: \citet{simon2011regularization} describe and implement an impressively fast algorithm \algorithm{coxnet}, implemented in the \pkg{glmnet} package, for fitting regularized \CPH models via weighted cyclic coordinate descent. This method is computationally efficient in handling high-dimensional problems. \citet{yang2013cocktail} proposed and implemented the \algorithm{cocktail} algorithm, which is a mixture of coordinate descent, the majorization-minimization principle, and the strong rule for solving penalized \CPH models in high dimensional data. The \algorithm{cocktail} algorithm (implemented in the \pkg{fastcox} package) always converges to the correct solution and is slightly faster than the \algorithm{coxnet} algorithm. However, these implementations, the benchmark \proglang{R} packages for penalized Cox models, have some limitations. The implementations by \cite{simon2011regularization} and \cite{yang2013cocktail} do not support time-dependent covariates; the implementation by \cite{goeman2010l1} does incorporate time-dependent covariates, but only implements \emph{naive} elastic net, neglecting subsequent improvements in the algorithm \citep{simon2011regularization}.

Other, non-\CPH-based, approaches have also incorporated time-dependent covariates in penalized models for time-to-event-data. Most such approaches have used generalized additive models to implement semiparametric regression methods in the context of survival models  \citep{gorst2012coordinate, bender2018generalized}. \cite{gorst2012coordinate} used a cyclic coordinate descent algorithm to develop  a penalized semiparametric additive hazard model (in the \pkg{ahaz} package). The model defines a hazard function as the sum of the baseline hazard and the regression function of the covariates --- it is intrinsically linear thus theoretically guarantees convergence, and can handle time-dependent covariates. However, currently, it only implements lasso penalization.

In this paper, we describe and implement an algorithm and \proglang{R} package (\pkg{pcoxtime}) for penalized \CPH models with time-dependent covariates. The general properties of penalized methods make this algorithm a useful tool for handling  high-dimensional problems. We describe how existing computational approaches for \CPH modeling can be adapted to obtain penalized methods for time-dependent covariates in time-to-event data. To solve the optimization problem, we exploit a variant of the gradient descent algorithm known as proximal gradient descent (as outlined in \cite{parikh2014proximal}) with Barzilai-Borwein step-size adjustment \citep{barzilai1988two}.
Unfortunately, the gradient-descent approach here is intrinsically slower than methods based on coordinate descent \citep{simon2011regularization};
we are working to implement coordinate descent. In the meantime, the capabilities and convenience of \pkg{pcoxtime} will still be useful  for moderately large problems.

We test our package on simulated data with time-dependent covariates, and compare its performance with that of the \pkg{penalized}. 
We also provide examples of its usage on real data. 

\section{Methods and algorithms} \label{sec:methods_algorithm}

\subsection{Cox model with time-independent covariates}\label{subsec:cox_independ}

Survival data is often presented in the form  $\{t_i, \delta_i, x_i\}^{n}_{i=1}$, where $t_i$ is the observed event time (failure time or censoring time) for individual $i$, $\delta_i$ is an indicator variable for whether the observed endpoint is a failure (rather than censoring), and $x_i$ is  a vector of covariates $(x_{i,1}, x_{i,2}, \cdots, x_{i,p})$. 

The \CPH model \citep{cox1972regression} defines the hazard function at time $t$  as 
\begin{equation} \label{eq:cox_hazard}
h_i(t) = h_0(t)\exp{(x_i^T\beta)},
\end{equation}
where $h_{0}(t)$ is the non-parametric baseline hazard and $\beta$ is the coefficient vector of length $p$.

In a simple case where there are no ties, with $t_1 < t_2 < \cdots < t_k$ representing unique ordered event (or failure) times, we can define the \emph{risk set} $R_i$, of individuals who are still at risk of failing (not yet censored or failed) at time $t_i$ -- individuals with event time $t_j\geq t_i$. The likelihood function corresponding to the order of events \citep{simon2011regularization, yang2013cocktail} is given by 
\begin{align}
\mathit{L(\beta)} = \prod_{i:\delta_i=1}{\frac{\exp{(x_i^T\beta)}}{\sum_{j\in R_i}\exp{(x_j^T\beta)}}} \label{eq:cox_hazard_likelihood},
\end{align}
and we can thus optimize the parameters $\beta$ by maximizing the partial log-likelihood: 
\begin{align}\label{eq:cox_hazard_loglikelihood}
\ell(\beta) = \sum_{i:\delta_i=1}\left(x_i^T\beta - \log\left[ \sum_{j\in R_i} \exp{(x_j^T\beta)} \right]\right).
\end{align}
The Cox model in \autoref{eq:cox_hazard} is fitted in two steps --- first, the parametric part is fitted by maximizing the partial log-likelihood in \autoref{eq:cox_hazard_loglikelihood}, and then the non-parametric baseline hazard is estimated.

Following the \pkg{survival} \citep{survival-package} package, here we use a slightly more general formulation, where the observed survival data is of the form $\{t^{\sstart}_{i}, t^{\sstop}_{i}, \delta_i, x_i\}^{n}_{i=1}$, where $t^{\sstart}_{i}$ and $t^{\sstop}_{i}$ bracket the period in which the event time for the $i$\textsuperscript{th} individual occurred. This formulation allows for greater flexibility in defining the time scale on which the analysis is based (e.g., time since diagnosis vs.~time of followup); it will also allow us to address left censorship in the observation of outcomes and (later) in the observation of covariates. The risk set at time $t_i$ is now defined as $R_i(t) = \{j : (t^{\sstart}_{j} < t_i) \cap (t_i \leq t^{\sstop}_{j})\}$. The first condition, $(t^{\sstart}_{j} < t_i)$, ensures the start time was observed before the event, while the second condition, $(t_i \leq t^{\sstop}_{j})$, ensures that individual $j$ either experienced the event or was censored at a later time point than $t_i$.

\subsection{Cox model with time-dependent covariates}\label{subsec:cox_depend}

When a covariate changes over time during the follow-up period, the observed survival data is of the form $\{t^{\sstart}_{i}, t^{\sstop}_{i}, \delta_i, x_i(t)\}^{n}_{i=1}$. The only difference is that $x_i$ is now a (piecewise constant) function of time. Using Breslow's approximation \citep{breslow1972contribution} for tied events, the partial log-likelihood is defined as 
\begin{align}\label{eq:cox_depend_loglike}
\ell(\beta) &= \sum_{i=1}^k \Bigg(\Bigg[\sum_{s\in D_i} x_s^T(t) \beta \Bigg]  - d_i\log \Bigg[\sum_{j\in R_i(t)} \exp(x_j^T(t) \beta) \Bigg] \Bigg),
\end{align}
where $d_i$ is the number of failures at time $t_i$, $k < n$ (if there are ties), $D_i$ are the set of indexes $j$ for subjects failing at time $t_i$ and the description of $t_1, t_2, \cdots, t_k$ remains the same as in the previous case \citep{harrell2015regression}. The parameter estimates $\hat{\beta}$ are obtained by minimizing $-\ell(\beta)$.

\subsection{Algorithm}\label{subsec:penalized_cox_algorithm}

Our algorithm extends the partial log-likelihood in \autoref{eq:cox_depend_loglike} by adding the penalty term. We let $P_{\alpha, \lambda}(\beta)$ be a mixture of $\ell_1$ (lasso) and $\ell_2$ (ridge) penalties. The penalized Cox partial log-likelihood (objective function) is defined as
\begin{equation*}
\penL(\beta)_{\alpha, \lambda} = -\mathit{\ell(\beta)} + P_{\alpha, \lambda}(\beta),
\end{equation*}
where
\begin{equation}
P_{\alpha, \lambda}(\beta) = \lambda\left(\alpha\sum_{i=1}^p|\beta_i| + 0.5(1 - \alpha)\sum_{i=1}^p\beta_i^2 \right) \label{eq:elasticnet1}
\end{equation}
as proposed by \cite{zou2005regularization}, with $\lambda > 0$ and $0 \leq \alpha\leq 1$. The lasso penalty ($\sum_{i=1}^p|\beta_i|$) induces sparsity by selecting a subset of nonzero coefficients.  It works well in high-dimensional  applications, but will eliminate all but one of any set of strongly multicollinear terms. On the other hand, the ridge penalty ($\sum_{i=1}^p\beta_i^2$) shrinks coefficients towards but never all the way to zero; hence it gives non-sparse estimates and can give correlated predictors approximately equal weights. The elastic net penalty combines the strength of lasso and ridge penalties for improved predictive performance \citep{simon2011regularization}. As $\alpha$ increases, the sparsity and the magnitude of non-zero coefficients decreases, i.e., the solution becomes less ridge-like and more lasso-like.

Using \autoref{eq:elasticnet1}, our minimization problem becomes
\begin{equation}\label{eq:elnet_loss}
\hat{\beta} = \arg \min_{\beta} \hspace{2mm} \penL(\beta)_{\alpha. \lambda}.
\end{equation}

\subsubsection{Parameter estimation}

The lasso penalty is not differentiable at $\beta = 0$. We thus solve the minimization problem above using \emph{proximal gradient descent} by decomposing the objective function (\autoref{eq:elnet_loss}) as $f(\beta) = g(\beta) + h(\beta)$, with
\begin{align*}
g(\beta) &= -\ell(\beta) + 0.5\lambda(1 - \alpha)\sum_{i=1}^p\beta_i^2
\end{align*}
and
\begin{align*}
h(\beta) &= \lambda\alpha\sum_{i=1}^p|\beta_i|.
\end{align*}
In this form, we split the objective function, $\arg \min_{\beta} \hspace{2mm} \penL(\beta)_{\alpha, \lambda}$, into two parts, one of which is differentiable. Specifically, $g(\beta)$ is differentiable and convex and $h(\beta)$ is convex but not necessarily differentiable. The proximal gradient operator \citep{parikh2014proximal} to update $\beta$ is given by
\begin{equation}\label{eq:prox_update}
\beta^{(k)} = \mathrm{prox}_{\gamma_{k}h}\left(\beta^{(k-1)} - \gamma_k\nabla g(\beta^{(k-1)})\right), ~ k = 1, 2, 3, \cdots
\end{equation}
where $\gamma_{k}$ is the step size determined via Barzilai-Borwein step-size adjustment \citep{barzilai1988two}. \cite{park2007l1} shows that  $\mathrm{prox}_{\gamma_{k}h}(.)$ reduces to (elementwise) \emph{soft thresholding}
\begin{align}
\mathrm{prox}_{\gamma_{k}\lambda\alpha}(x_i) = 
\begin{cases}
  x_i - \gamma_k\lambda\alpha & x_i \geq \gamma_k\lambda\alpha\\    
  0 & -\gamma_k\lambda\alpha\leq x_i \leq \gamma_k\lambda\alpha\\
  x_i + \gamma_k\lambda\alpha & x_i \leq -\gamma_k\lambda\alpha 
\end{cases} \label{eqn:thresholding}
\end{align}
and
\begin{align}
\nabla g(\beta) &= -\nabla\ell(\beta) + \lambda(1 - \alpha)\sum_{i=1}^p\beta_i\nonumber\\
&= - \sum_{i=1}^k \Bigg(\sum_{s\in D_i} x_s^T(t)  - d_i \frac{\sum_{j\in R_i(t)} x_j^T(t) \exp(x_j^T(t) \beta)}{\sum_{j\in R_i(t)} \exp(x_j^T(t) \beta)} \Bigg) + \lambda(1 - \alpha)\sum_{i=1}^p\beta_i \nonumber \\
&= \pi(\beta) + \lambda(1 - \alpha)\sum_{i=1}^p\beta_i. \label{eqn:loglik_derivative}
\end{align}
Our package implements the Karush–Kuhn–Tucker (KKT) conditions check described in \cite{yang2013cocktail} to test that the $\beta$  estimates are valid. We did not come across any convergence problems in the examples analyzed here (i.e., the KKT conditions were always satisfied)


To train an optimal model, we need to choose a value of $\lambda$. With a large value of $\lambda$ the penalty terms in \autoref{eq:elnet_loss} will dominate, driving coefficients to zero, while a small $\lambda$ value will lead to overfitting. We can use cross-validation to pick an optimal $\lambda$ from a set of $\lambda$ values (known as a regularization path) $\lambda_1 < \lambda_2, \cdots, <\lammx$. We want \lammx\ to be large enough that $\beta = \boldsymbol 0$, and $\lambda_1$ to be small enough to give a result close to the unpenalized solution (this choice enables the \textit{warm-start} approach employed in \pkg{glmnet}). 

From \autoref{eqn:thresholding} and \autoref{eqn:loglik_derivative} notice that if $\pi(\beta) \leq \alpha\lambda\gamma_k$, then $\beta^k = 0$ minimizes our objective function. Thus we set $\lammx$ to be
\begin{equation}\label{eq:lambdamax}
\lambda_{\mathrm{max}} = \frac{1}{N\alpha\gamma_k}\max\limits_{\beta}\left\lbrace\sum_{i=1}^k \Bigg(\Bigg[\sum_{s\in D_i} x_s^T(t) \Bigg]  - \frac{d_i}{|R_i(t)|} \Bigg[\sum_{j\in R_i(t)} x_j^T(t) \Bigg] \Bigg)\right\rbrace,
\end{equation}
where $|R_i(t)|$ denotes the cardinality of the risk set $R_i(t)$. If $\alpha = 0$, we set $1/{N\alpha\gamma_k}$ in \autoref{eq:lambdamax} to $1/{0.001N\gamma_k}$.

In our implementation, we set $\lambda_{\mathrm{min}} = \epsilon \lambda_{\mathrm{max}}$, and compute solutions over a grid of $m$ values of $\lambda$ decreasing from $\lambda_{\mathrm{max}}$ to $\lambda_{\mathrm{min}}$, where $\lambda_{i} = \lambda_{\max}(\lambda_{\min}/\lambda_{\max})^{i/(m-1)}$ for $i = 0,\cdots, m-1$ \citep{simon2011regularization}. The default value of $k$ is $100$ (the number of distinct $\lambda$ values). If $n \geq p$, the default value of $\epsilon$ is set to $0.0001$; otherwise (i.e. if $n < p$), $\epsilon = 0.01$ \citep{yang2013cocktail, simon2011regularization}.

\subsubsection{Cross-validation}

Most implementations cross-validate over a range of $\lambda$ values for a fixed $\alpha$. However, our implementation allows the user to choose a range of $\alpha$, $0 \leq \alpha\leq 1$; in this case the algorithm will pick the $\alpha$-$\lambda$ pair that corresponds to the lowest cross-validated partial likelihood deviance (CV-PLD) or highest cross-validated Harrell's concordance index (CV-C-index) \citep{harrell1996multivariable}.

To find the CV-PLD for each $\lambda$-$\alpha$ pair, we perform $k$-fold cross-validation --- the training data is split into $k$ folds, and the model is trained on $k-1$ folds and validated on the left-out part via some predictive performance measure $k$ times. Here, we implement two metrics i.e., CV-PLD and  CV-C-index \citep{dai2019cross}. The CV-PLD is:
\begin{equation}\label{eq:cve}
\CVPL = -2\sum_{k=1}^K{\ell(\hat{\beta}_{-k}(\lambda)) - \ell_{-k}(\hat{\beta}_{-k}(\lambda))}
\end{equation}
where $\ell(\hat{\beta}_{-k}(\lambda))$ is the log partial likelihood evaluated at $\hat{\beta}_{-k}$ using the whole dataset and $\ell_{-k}(\hat{\beta}_{-k}(\lambda))$ is the log partial likelihood evaluated at  $\hat{\beta}_{-k}$ on the retained data (everything except the left-out part). The $\hat{\beta}_{-k}$ values denote the penalized estimates using the retained data. We choose the $\lambda$ which minimizes \autoref{eq:cve}. Note that \autoref{eq:cve} gives different (and usually better) results than simply evaluating the partial likelihood on the held-out set (sometimes called the \emph{basic} approach), because the likelihood of any observation depends on other elements in the risk set. 

The alternative cross-validation metric, CV-C-index, uses the concordance statistic for Cox models, known as the cross-validated $C$-index, based on Harrell's concordance index \citep{harrell1996multivariable}. It computes the probability that, for a random pair of individuals, the predicted survival times of the pair have the same ordering as their true survival times. Our implementation is similar to that of the \pkg{survival} package \citep{survival-package}.

\subsection{Prediction}
Once $\hat{\beta}$ is estimated, we can estimate the baseline hazard function ($\hat{h}_0(t)$), and hence the survival function ($\hat{S}_0(t)$). We first compute the cumulative hazard function 
\begin{equation}\label{eq:estimated_basehaz}
\hat{h}_0(t) = \sum_{i\in y_j < t_i}{\frac{\sigma_i}{\sum_{j\in R_i(t)}\exp(x(t)_j^T\hat{\beta})}}
\end{equation}
and then, for a given covariate vector, $x_i$, the estimated hazard, $\hat{h}(t|x_i)_i$, and survival functions are
\begin{align*}
\hat{h}(t|x_i)_i &= \hat{h}_0(t)\exp{(x_i^T(t)\hat{\beta})}\\
\hat{S}(t|x_i)_i &= \exp{\left(-\hat{h}(t|x_i(t))_i\exp{(x_i^T(t)\hat{\beta})}\right)}.
\end{align*}
%


\section{Illustrations} \label{sec:illustrations}

In the following sections, we demonstrate the practical use of \pkg{pcoxtime} on real and simulated data sets. In the first two examples, we consider real data sets with time-independent covariates and then a time-dependent covariates. The last example compares \pkg{pcoxtime} with \pkg{penalized} on a simulated data set with time-dependent covariates.

\subsection{Time-independent covariates}

We use the \code{sorlie} gene expression data set \citep{sorlie2003parker}, which contains $549$ gene expression measurements together with the survival times for $115$ females diagnosed with cancer. This data set was also used by \cite{gorst2012coordinate} to demonstrate the performance of the \pkg{ahaz} package.

We perform a penalized survival regression by varying both $\alpha$ and  $\lambda$. If a range of $\alpha$ values is desired, we suggest first running the analysis for an intermediate range of $\alpha$ values. If the minimum cross-validation likelihood deviance (\minCVPL) based on $k$-fold cross-validation (over all $\lambda$ values considered) is at the lower bound of the range considered, then extend the range of the $\alpha$ vector to lower (positive) values; if it is at the upper bound, extend the range upward. In this example, we cross-validate several $\alpha$ values at the same time by setting $\alpha = \{0.1, 0.2, 0.4, 0.6, 0.8, 1\}$. For each $\alpha$, we analyze a solution path  of $\lambda$ values and use 10-fold cross validation to choose the  optimal (\minCVPL) value of $\alpha$ and $\lambda$.

We first load the data as follows:

\begin{CodeChunk}
\begin{CodeInput}
R> data("sorlie", package = "ahaz")
\end{CodeInput}
\end{CodeChunk}

It is common practice to standardize the predictors before applying penalized methods.  In \pkg{pcoxtime}, predictors are scaled internally (but the user can choose to output  coefficients on the original scale [the default] or to output standardized coefficients). We make the following call to \algorithm{pcoxtimecv} to perform 10-fold cross-validation to choose the optimal $\alpha$ and $\lambda$.

\begin{CodeChunk}
\begin{CodeInput}
R> cv_fit1 <- pcoxtimecv(Surv(time, status) ~., data = sorlie, 
+    alphas = c(0.1, 0.2, 0.4, 0.6, 0.8, 1), lambdas = NULL, 
+    devtype = "vv", lamfract = 0.8, refit = TRUE, nclusters = 4
+  )
Progress: Refitting with optimal lambdas...  
\end{CodeInput}
\end{CodeChunk}

In order to reduce the computation time, we use \code{lamfract} to set the proportion of $\lambda$ values, starting from $\lammx$, to $80\%$. Setting \code{lamfract} in this way specifies that only a subset of the full sequence of $\lambda$ values is used \citep{simon2011regularization}.

\begin{figure}[!t]
\begin{center}
\includegraphics{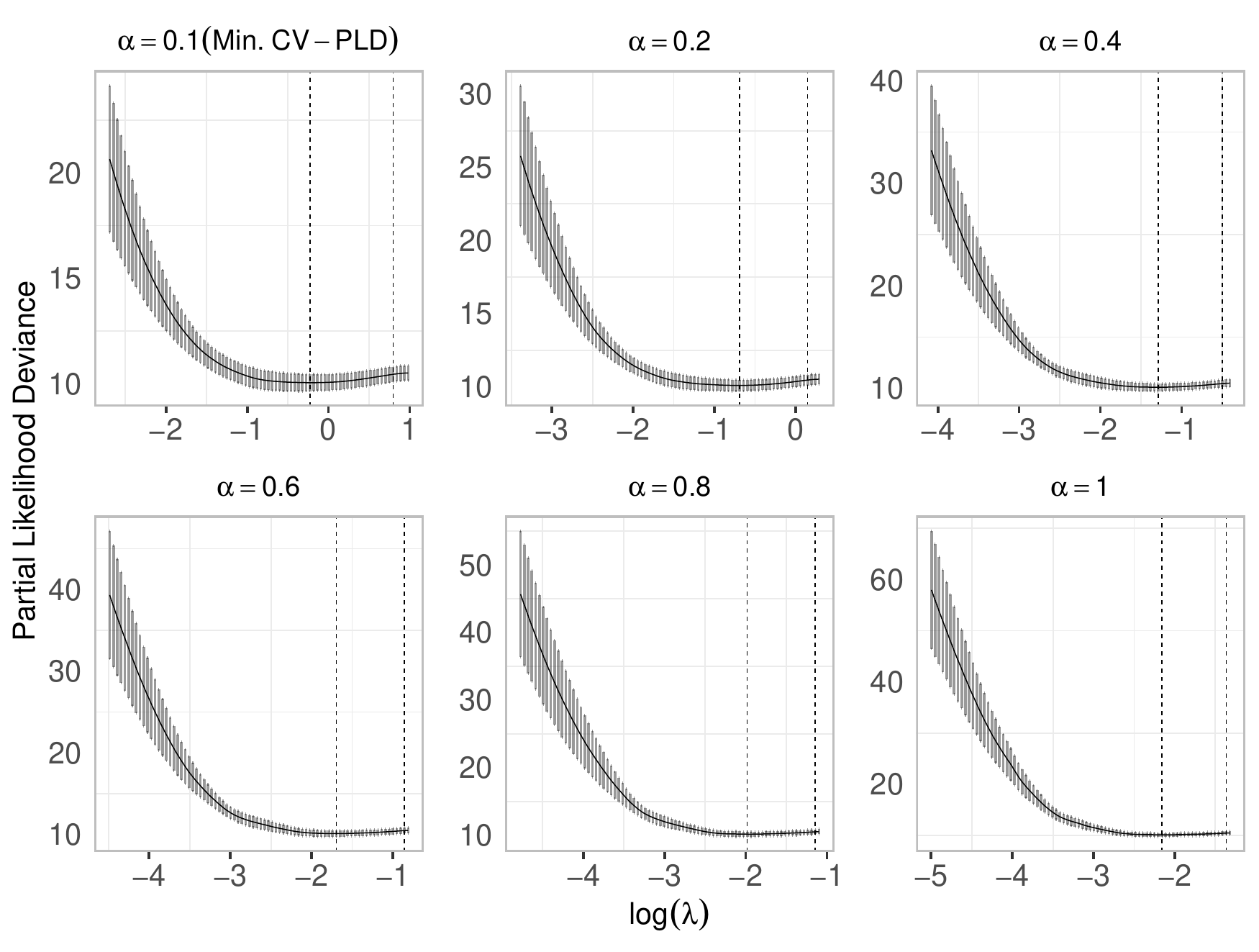}
\end{center}
\caption{\label{fig:error_rates_sorlie} Plots of the cross-validated error rates for a sequence of $\lambda$ values at different $\alpha$ values. If a vector of $\alpha$ values is specified, \algorithm{pcoxtimecv} automatically chooses the $\alpha$ that minimizes CV-PLD (\minCVPL). In this example, we choose from $\alpha = \{0.1, 0.2, 0.4, 0.6, 0.8, 1\}$. The left dotted line indicates the minimum error; the right dotted line indicates the largest value of $\lambda$ that fits the simplest model whose error is within one standard deviation of the minimum cross-validation error \citep{hastie2009elements}. In this case, the best fit occurs when $\alpha = 0.1$ (more ridge-like, top left panel); note that panels differ in their horizontal and vertical scales.
}
\end{figure}

Once cross-validation is performed, we can report the $\lambda$ and $\alpha$ for which CVE attains its minimum ($\lambda=0.796, \alpha=0.1$) and view the cross-validated error plot (\autoref{fig:error_rates_sorlie}) and the regularization path (\autoref{fig:solution_path_sorlie_cgd}).

\begin{CodeChunk}
\begin{CodeInput}
R> print(cv_fit1)
Call:
pcoxtimecv(formula = Surv(time, status) ~ ., data = sorlie, alphas = c(0, 
    1), lambdas = NULL, lamfract = 0.8, devtype = "vv", refit = TRUE, 
    nclusters = 4)

Optimal parameter values
 lambda.min lambda.1se alpha.optimal
  0.7960954   2.215183           0.1
R>
R> cv_error1 <- plot(cv_fit1, g.col = "black", geom = "line", 
+     scales = "free")
R> print(cv_error1)
R>
R> solution_path1 <- (plot(cv_fit1, type = "fit") + 
+    #ylim(c(-0.03, 0.03)) + 
+    labs(caption = "(a) sorlie") +
+    theme(plot.caption = element_text(hjust=0.5, size=rel(1.2)))
+ )
R> print(solution_path1)
\end{CodeInput}
\end{CodeChunk}

Next, we fit the penalized model using the optimal $\alpha$ and $\lambda$:

\begin{CodeChunk}
\begin{CodeInput}
R> ## Optimal lambda and alpha
R> alp <- cv_fit1$alpha.optimal
R> lam <- cv_fit1$lambda.min
R> 
R> ## Fit penalized cox model
R> fit1 <- pcoxtime(Surv(time, status) ~., data = sorlie,
+    alpha = alp, lambda = lam
+  )
R> print(fit1)
Call:
pcoxtime(formula = Surv(time, status) ~ ., data = sorlie, alpha = alp, 
    lambda = lam)

66 out of 549 coefficients are nonzero
n = 115 , number of events = 38 
\end{CodeInput}
\end{CodeChunk}

We then plot the predicted survival function for each patient and the average survival function (\autoref{fig:survival_curve_sorlie_cgd}).
\begin{CodeChunk}
\begin{CodeInput}
R> surv_avg <- pcoxsurvfit(fit1)
R> surv_df <- with(surv_avg, data.frame(time, surv))
R> surv_ind <- pcoxsurvfit(fit1, newdata = sorlie)
R> splot_sorlie <- plot(surv_ind, lsize = 0.05, lcol="grey")
R> splot_sorlie <- (splot_sorlie + 
+    geom_line(data = surv_df, aes(x = time, y = surv, group = 1),
+    col = "red") +
+    labs(caption = "(a) sorlie") + 
+    theme(plot.caption = element_text(hjust=0.5, size=rel(1.2)))
+ )
R> print(splot_sorlie)
\end{CodeInput}
\end{CodeChunk}

\subsection{Time-dependent covariates}\label{subsec:time-dep}

We now repeat the analysis outlined above in the context of survival data with time-dependent covariates. We consider the chronic granulotomous disease (\code{cgd}) data set from the \pkg{survival} package \citep{survival-package}, which contains data on time to serious  infection for 128 unique patients. Because some patients are observed for more than one time interval, with different covariates in each interval, the data set has 203 total observations.

\begin{figure}[!t]
\centering
\includegraphics{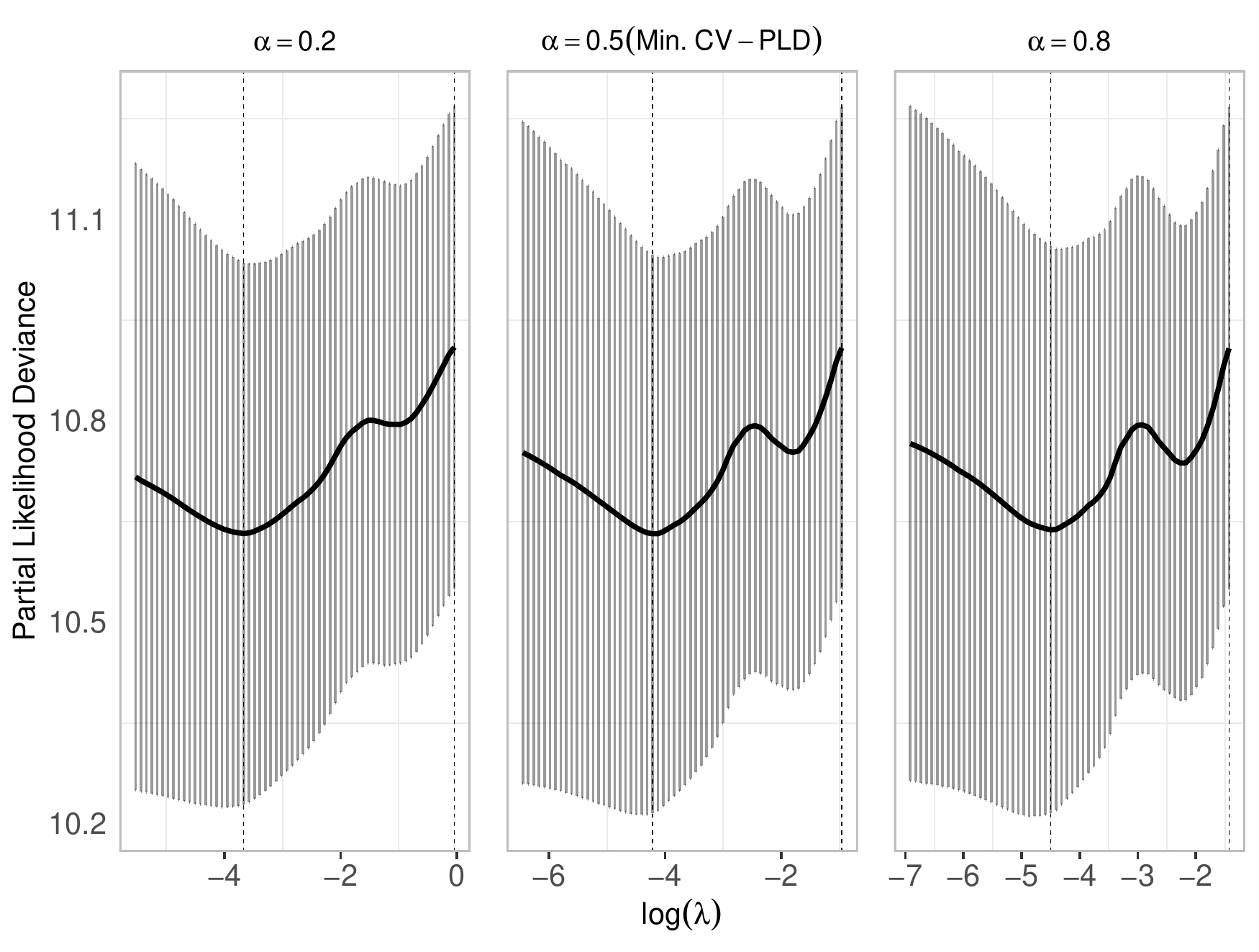}
\caption{\label{fig:error_rates_cgd} Cross-validated error rates for the \code{cgd} data.}
\end{figure}

We load the data and perform cross-validation:

\begin{CodeChunk}
\begin{CodeInput}
R> data("cgd", package = "survival")
R> dat <- cgd
R> cv_fit2 <- pcoxtimecv(Surv(tstart, tstop, status) ~ treat + sex + 
+    ns(age,3) + height + weight + inherit + steroids + propylac + 
+    hos.cat, data = cgd, alphas = c(0.2, 0.5, 0.8), lambdas = NULL, 
+    devtype = "vv", lamfract = 0.6,  refit = TRUE, nclusters = 4
+  )
Progress: Refitting with optimal lambdas... 
\end{CodeInput}
\end{CodeChunk}
Here, we choose $\alpha = \{0.2, 0.5, 0.8\}$ for $10$-fold cross-validation. 
\begin{CodeChunk}
\begin{CodeInput}
R> print(cv_fit2)
Call:
pcoxtimecv(formula = Surv(tstart, tstop, status) ~ treat + sex + 
    ns(age, 3) + height + weight + inherit + steroids + propylac + 
    hos.cat, data = cgd, alphas = c(0.2, 0.5, 0.8), lambdas = NULL, 
    lamfract = 0.6, devtype = "vv", refit = TRUE, nclusters = 4)

Optimal parameter values
 lambda.min lambda.1se alpha.optimal
 0.01477729  0.3834742           0.5
\end{CodeInput}
\end{CodeChunk}
The \minCVPL (optimal) hyperparameter values are $\lambda=0.015$ and $\alpha=0.5$ (Figure \autoref{fig:error_rates_cgd}).
\begin{CodeChunk}
\begin{CodeInput}
R> cv_error2 <- plot(cv_fit2, g.col = "black", geom = "line", 
+     g.size = 1)
R> print(cv_error2)
\end{CodeInput}
\end{CodeChunk}

\begin{figure}[!t]
\centering
\includegraphics{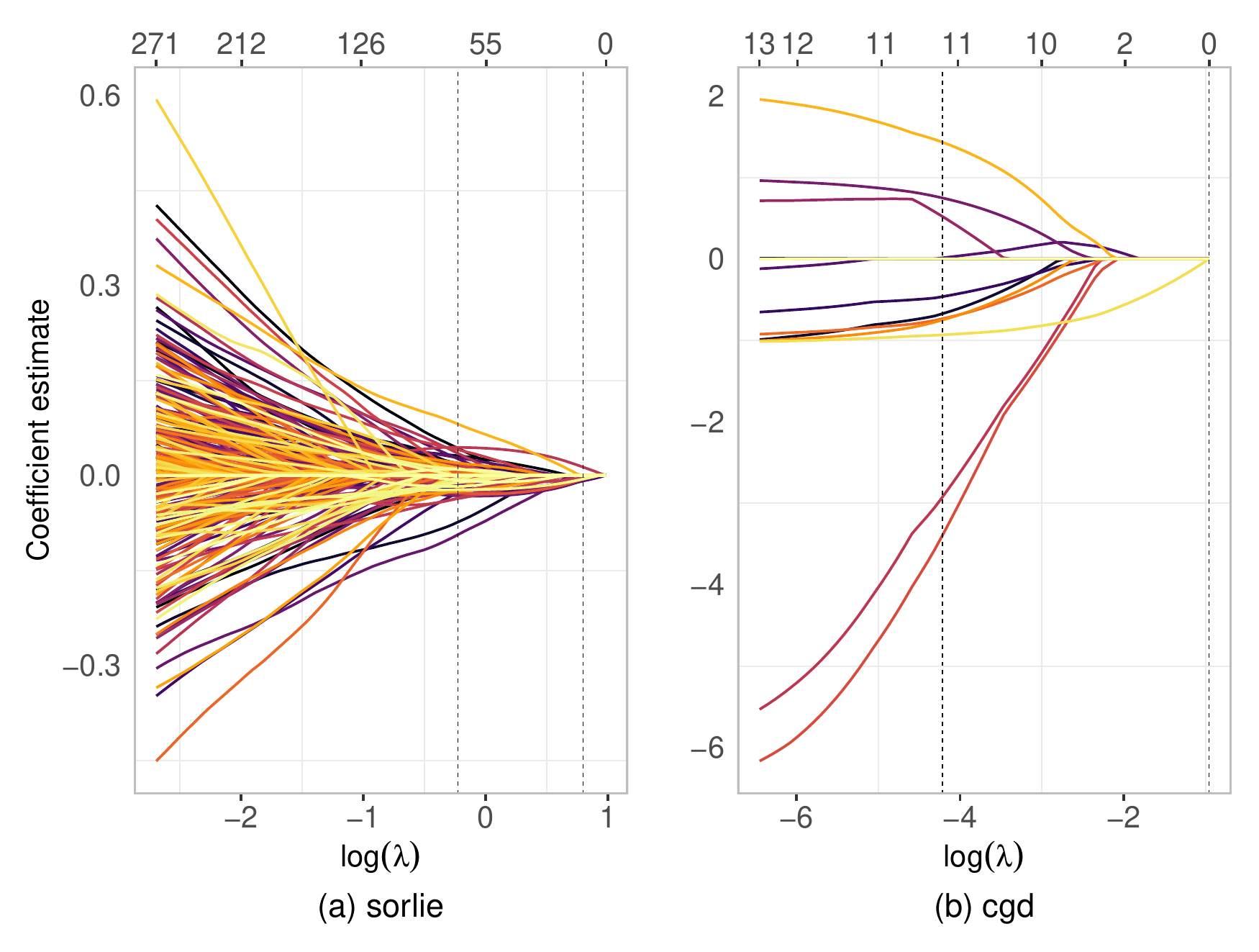}
\caption{\label{fig:solution_path_sorlie_cgd} Regularization paths for the \code{sorlie} and \code{cgd} models. The values at the top of the plot give the number of nonzero coefficients (size of the model) at various $\lambda$ values.}
\end{figure}

We plot the solution path (\autoref{fig:solution_path_sorlie_cgd}) and fit the penalized model based on the optimal $\lambda$ and $\alpha$:
\begin{CodeChunk}
\begin{CodeInput}
R> solution_path2 <- (plot(cv_fit2, type = "fit") + 
+    labs(caption = "(b) cgd") + 
+    theme(plot.caption = element_text(hjust=0.5, size=rel(1.2)))
+ )
R> print(solution_path2)
R>
R> alp <- cv_fit2$alpha.optimal
R> lam <- cv_fit2$lambda.min
R> 
R> ## Fit penalized cox model
R> fit2 <- pcoxtime(Surv(tstart, tstop, status) ~ treat + sex + 
+    ns(age,3) + height + weight + inherit + steroids + propylac + 
+    hos.cat, data = cgd, alpha = alp, lambda = lam
+  )
R> print(fit2)
Call:
pcoxtime(formula = Surv(tstart, tstop, status) ~ treat + sex + 
    ns(age, 3) + height + weight + inherit + steroids + propylac + 
    hos.cat, data = cgd, alpha = alp, lambda = lam)

11 out of 13 coefficients are nonzero
n = 203, number of events = 76 
\end{CodeInput}
\end{CodeChunk}

\begin{figure}[!t]
\centering
\includegraphics{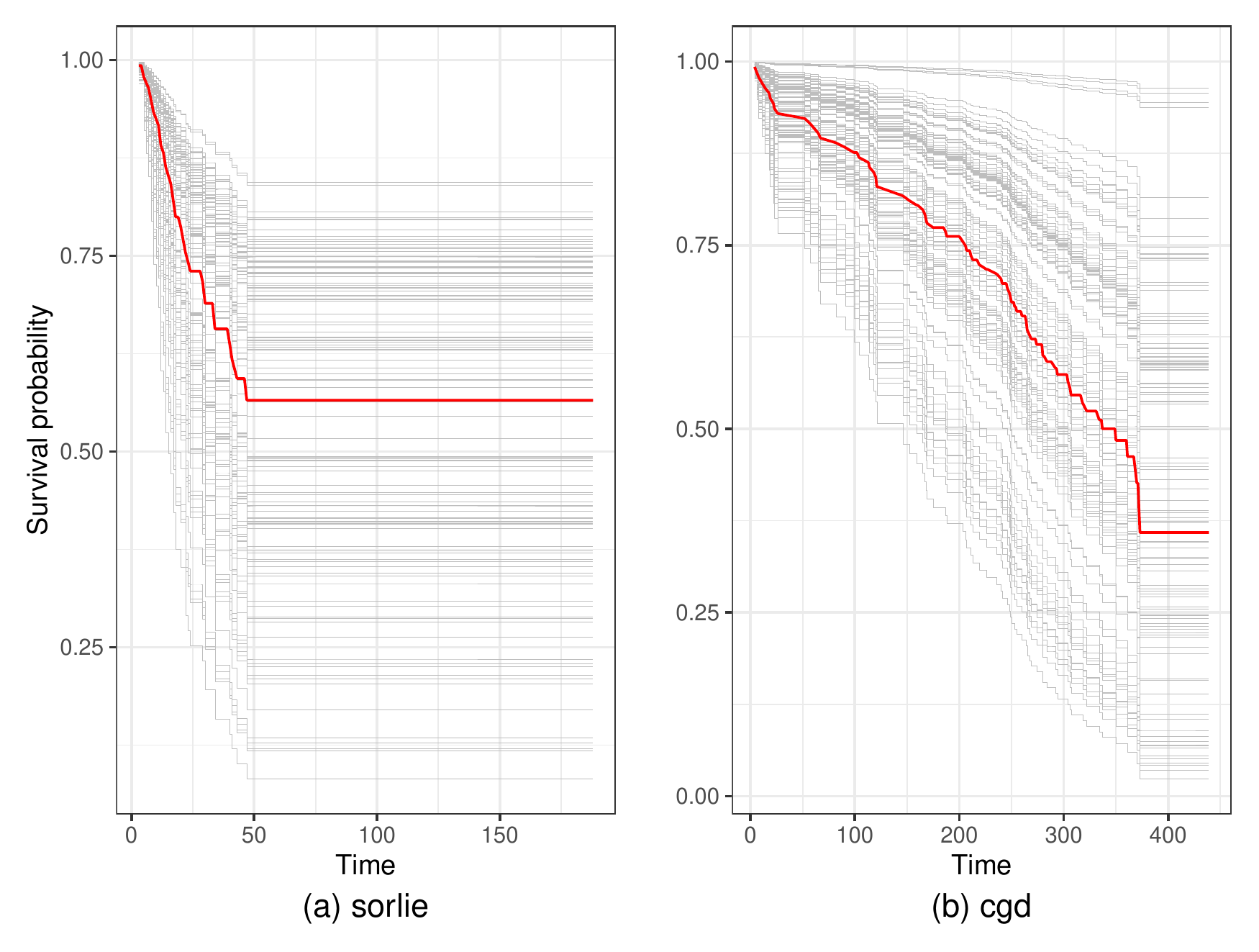}
\caption{\label{fig:survival_curve_sorlie_cgd} Predicted individual and average (\textcolor{red}{red}) survival probabilities for \code{sorlie}, \emph{left}, and \code{cgd}, \emph{right}, data sets.}
\end{figure}

Again, we use the \code{fit2} object to  plot the predicted individual and average survival curves (\autoref{fig:survival_curve_sorlie_cgd}). 
\begin{CodeChunk}
\begin{CodeInput}
R> surv_avg <- pcoxsurvfit(fit2)
R> surv_df <- with(surv_avg, data.frame(time, surv))
R> surv_ind <- pcoxsurvfit(fit2, newdata = cgd)
R> splot_cgd <- (plot(surv_ind, lsize = 0.05, lcol = "grey") + 
+    geom_line(data = surv_df, aes(x = time, y = surv, group = 1),
+    col = "red") + 
+    labs(caption = "(b) cgd") + 
+    theme(plot.caption = element_text(hjust=0.5, size=rel(1.2)))
+ )
R> print(splot_cgd)
\end{CodeInput}
\end{CodeChunk}

\subsection{Simulated data set}\label{ssec:simdata}

In this section, we test our package on simulated data with time-dependent covariates, and compare its performance with that of the \pkg{penalized} algorithm. We first describe our data simulation process and then report the performance results.

We provide a user-friendly wrapper, \code{simtdc}, for the extended permutational algorithm for simulating time-dependent covariates provided by the \pkg{PermAlgo} package
\citep{sylvestre2008comparison}.

We simulated a data set for 120 unique individuals with a follow-up time of up to $10$ time units (years), $100$ time-dependent and $900$ time-fixed covariates --- all drawn from a normal distribution with a mean of $0$ and standard deviation of $1$ with the true effect size, expressed on log hazard scale, of each covariate drawn from a uniform distribution [0, 2]. Since some individuals were observed in more than one time interval there were 444 observations, of which we used 299 observations for training and the remainder for testing. Covariates affected relative hazard only; event times were chosen assuming a constant total hazard rate of 0.2. Censoring times were chosen uniformly over the time period. 

We compared our proximal gradient descent algorithm, \pkg{pcoxtime}, to the combination gradient descent-Newton-Raphson method from \pkg{penalized} \citep{goeman2010l1}. The two packages use different elastic net penalty specifications (\pkg{penalized} uses $\lambda_1$ and $\lambda_2$ for the lasso and ridge penalties instead of an overall $\lambda$ and a mixing parameter $\alpha$). We used $\alpha = 0.5$ and the default range of $\lambda$ values in \pkg{pcoxtime}, then used a convenience function from [the development version of] \pkg{pcoxtime} to calculate values of $\lambda_1$ and $\lambda_2$ for \pkg{penalized}. Although the two approaches are similar, \pkg{penalized} has two possible limitations: (1) it cross-validates elastic net in two steps, finding a value of the ridge penalty $\lambda_2$ for each value of the lasso penalty $\lambda_1$ in order to fit an elastic net; for $k$-fold CV, this two-step procedure will require $k$ times as much computational effort. (2) Possibly to compensate for this inefficiency, it uses Brent's algorithm to search for the optimal hyperparameter values (rather than using a parameter grid as we and others do), which risks converging to a local optimum \citep{goeman2018package}. 

To compare the predictions of \pkg{penalized} and \pkg{pcoxtime}, we used two approaches to choose the hyperparameters for \pkg{penalized}: (1) ``\pkg{pcoxtime}-$\lambda_1$-based'', using the optimal $\alpha$ and  $\lambda$ chosen by \pcoxtime model to calculate the $\lambda_1$ and $\lambda_2$ values for the \pkg{penalized} model  (we call this the \emph{pcox-pen} model). (2) ``\pkg{penalized}-$\lambda_1$-based'', training the penalized model using the optimal $\lambda_1$ value determined by \pkg{penalized} from a vector of $\lambda$ values generated from \pcoxtime's cross-validation (we call this the \emph{pen-pen} model). The two predictions were then compared to \pcoxtime's estimates  (\emph{pcox}).

The two \pkg{pcoxtime} fits (\emph{pcox-pen} and \emph{pen-pen}) gave very similar estimates of the optimal $\lambda$ (16.31 and 14.94, respectively). Comparing these results with \pcoxtime's, all three  approaches gave similar estimates and confidence intervals for Harrell's $C$-statistic \citep{survival-package} (both \emph{pcox} and \emph{pcox-pen} gave $C=0.65 [0.51, 0.77]$,  while the \emph{pen-pen} values differed by a few percent: $C=0.64[0.50, 0.74]$).

The \pkg{pcoxtime} package took about 27 times as long as \pkg{penalized} to compute the complete solution path (1152 seconds vs. 42 seconds), probably because our current implementation uses \proglang{C++} only for likelihood computation and coefficient estimation for each $\lambda$; the solution paths are computed in \proglang{R}. All computations were carried out on an 1.80 GHz, 8 processors Intel Core i7 laptop. \autoref{tab:compare} compares the features of the different packages available for fitting penalized \CPH models.

\section{Comparison among packages}\label{sec:compare}

In this section, we compare the capabilities of \pkg{pcoxtime} to some of the most general and widely used \proglang{R} packages for penalized \CPH models --- \pkg{glmnet}, \pkg{fastcox} and \pkg{penalized}. Computational speed is important in high-dimensional data analysis; packages using coordinate descent based methods (\pkg{glmnet} and \pkg{fastcox}) are usually much faster than gradient descent based methods (\pkg{pcoxtime} and \pkg{penalized}) \citep{simon2011regularization}. \autoref{tab:compare} summarizes some important capabilities of the packages. 

\begin{table}[H]
\centering
\begin{tabular}{lllllllll}
\hline
           						&	& \pkg{pcoxtime} &	& \pkg{glmnet} &  & \pkg{fastcox} & 	& \pkg{penalized}\\ \hline
\emph{Supported models}     	&    &       		 &	&        		&  &			   &	& 					\\
Time-dependent covariates		&	&	yes			 &	&	no			& &	no				&   &	yes				\\
Penalty parameterization					&	& $\lambda$,	$\alpha$			 &	&	 $\lambda$,	$\alpha$			& &	 $\lambda$,	$\alpha$				&   &	 $\lambda_1$,	$\lambda_2$				\\ \hline
\emph{Post model predictions}     		&    &       		 &	&        		&  &			   &	& 					\\
Survival and hazard functions	&	&	yes			 &	&	no			& &	no				&   &	yes				\\
Model diagnostics \& validation  & & yes			 &	&	no			& &	no				&   &	no				\\
(prediction error, Brier score, && \\
calibration plots, etc.) \\
\hline
\end{tabular}
\caption{\label{tab:compare} Capabilities of \pkg{pcoxtime}, \pkg{glmnet}, \pkg{fastcox} and \pkg{penalized} packages.}
\end{table}

\section{Conclusion} \label{sec:summary}

We have shown how the penalized \CPH model can be extended to handle time-dependent covariates, using a proximal gradient descent algorithm. This paper provides a general overview of the \pkg{pcoxtime} package and serves as a starting point to further explore its capabilities.

In future, we plan to improve the functionality of \pkg{pcoxtime}. In particular, we plan to implement a coordinate descent algorithm in place of the current proximal gradient descent approach, which should greatly improve its speed.

\section*{Acknowledgments}

We would like to thank Mr. Erik Drysdale and Mr. Brian Kiprop for the discussions and valuable feedback on earlier versions of the package. This work was supported by a grant to Jonathan Dushoff from the Natural Sciences and Engineering Research Council of Canada (NSERC) Discovery.

\bibliography{refs}

\end{document}